\documentclass[aps,prx,twocolumn,amsmath,amssymb,superscriptaddress,floatfix]{revtex4-2}

\usepackage{mathtools}
\usepackage{multirow}
\usepackage{xcolor}
\usepackage{bm}
\usepackage{amsfonts,amssymb,amsmath}
\usepackage{graphicx,dcolumn,bm,xcolor,braket,slashed}
\usepackage{times} 
\usepackage{comment}
\usepackage{array}
\usepackage{textcomp}
\usepackage[normalem]{ulem}
\usepackage{dsfont}
\usepackage{kotex}

\newcommand{\be}{\begin{eqnarray}}
\newcommand{\ee}{\end{eqnarray}}
\newcommand{\bbm}{\begin{bmatrix}}
\newcommand{\ebm}{\end{bmatrix}}
\newcommand{\bpm}{\begin{pmatrix}}
\newcommand{\epm}{\end{pmatrix}}

\newcommand{\BZ}{\mathbb{T}^2}
\newcommand{\kk}{\bm{k}}

\newcommand{\Pbr}{P_{\mathrm{bi}}}

\newcommand{\PFB}{P_{\mathrm{FB}}}

\begin{document}

\title{ Classification of Non-Hermitian Flat Bands}

\author{Chang-geun Oh}
\email{cg.oh.0404@gmail.com}
\affiliation{Department of Applied Physics, The University of Tokyo, Tokyo 113-8656, Japan}
\author{Jun-Won Rhim}
\email{jwrhim@ajou.ac.kr}
\affiliation{Department of Physics, Ajou University, Suwon 16499, Republic of Korea}

\begin{abstract}
Exact flat bands provide a versatile setting for correlated and topological phenomena, yet their properties are controlled not only by their dispersion but also by the structure of their Bloch projectors. 
Here, we establish a projector-based classification of non-Hermitian flat bands. 
In contrast to Hermitian flat bands, non-Hermiticity introduces a biorthogonal projector whose left-right pairing permits a distinct norm-pole singularity. 
We identify four classes: analytic NH-A, continuous but nonanalytic NH-C, bounded but discontinuous NH-D, and pole-singular NH-E. 
We show that continuity of the biorthogonal projector preserves the locking of the right, left, and biorthogonal Chern numbers, $C_R=C_L=C_{\rm bi}$, thereby realizing non-Hermitian critical topological flat bands. 
Once continuity is lost, this locking can break down; in particular, a mismatch $C_R\neq C_L$ can occur in the NH-E class. 
Finally, we show that this Chern number mismatch ensures to produce an anomalous enhancement of resonant cross-orbital transfer. Our results establish biorthogonal projector regularity as the organizing principle linking compact localized states, topology, and driven response in non-Hermitian flat bands.
\end{abstract}

\maketitle

\section{Introduction}
\label{sec:intro}

Flat bands---electronic bands that remain exactly dispersionless across the entire Brillouin zone---have become a central arena for strongly correlated physics, since the complete quenching of the kinetic energy allows interactions to dominate the low-energy dynamics~\cite{LeykamReview,DerzhkoRev}. 
They underlie flat-band ferromagnetism~\cite{Mielke,Tasaki}, fractional Chern insulators~\cite{Tang2011,Sun2011,Neupert2011,Xie2021}, and the unconventional superconducting phases of magic-angle twisted bilayer graphene~\cite{CaoIns,CaoSC}.

The defining real-space hallmark of a flat band is the compact localized state (CLS), an eigenstate of strictly bounded support produced by destructive interference~\cite{Sutherland,Bergman2008}. 
In Hermitian systems, flat bands are fundamentally classified by a single momentum-space property: the regularity of the flat-band projector, as shown in Fig.~1. 
When the projector is analytic across the entire Brillouin zone (BZ), a prominent no-go theorem guarantees that the Chern number vanishes $C=0$~\cite{ChenMazaheriSeidelTang2014} and the CLS basis is complete, spanning the entire subband. 
Conversely, when the projector loses continuity at specific touching points, the Chern number becomes ill-defined and the CLS basis cannot span the entire flat band~\cite{RhimYang2019,RhimYang2021}. 
Flat bands in this class, termed singular flat bands, hosts rich physical phenomena including a quantum-distance-characterized bulk-interface correspondence~\cite{oh2022bulk,kim2023general} and anomalous Landau level spreading~\cite{rhimkimyang2020,RhimYang2021}. Bridging these limits, a continuous but non-analytic projector defines critical topological flat bands~\cite{li2026stable,yang2025fractional}, where the CLS basis remains incomplete yet a well-defined, potentially non-zero integer Chern number survives. 
A natural question follows: how does this geometric classification paradigm transform once Hermiticity is relaxed?

Non-Hermitian (NH) Hamiltonians describe open, dissipative, and gain--loss systems, hosting topological phenomena absent in the Hermitian realm~\cite{AshidaGongUeda2020,bergholtz2021exceptional,okuma2023non}. 
Because $H \neq H^\dagger$, the left and right eigenvectors of a band no longer coincide. 
Consequently, a single isolated subband features three distinct gauge-invariant projectors and, correspondingly, three Chern numbers: right ($C_R$), left ($C_L$), and biorthogonal ($C_{\rm bi}$). For a completely separable band, these invariants are inherently locked together, $C_R=C_L=C_{\rm bi}$, reducing the topology to a single integer~\cite{ShenZhenFu2018}.
The microscopic mechanism driving this synchronization is that a non-vanishing biorthogonal pairing, $\langle u^L(\bm{k})\vert u^R(\bm{k})\rangle \neq 0$, rigidly couples the right and left gauge choices~\cite{ShenZhenFu2018,KunstEdvardsson2018}. However, flat bands generically can touch with dispersive bands, meaning this gauge-locking argument is fundamentally challenged at the touching points where the pairing vanishes. 
Does the topological locking $C_R=C_L=C_{\rm bi}$ survive when a flat band undergoes a band touching? If the locking fails, what physical signatures distinguish the resulting geometric regimes?

In this Letter, we resolve these questions through a projector-based classification of NH flat bands. 
The central object is the biorthogonal flat-band projector $\Pbr(\bm{k})$. 
The regularity of $\Pbr(\bm{k})$ systematically sorts every flat band into four distinct classes: analytic (NH-A), continuous but non-analytic (NH-C), bounded but discontinuous (NH-D), and singular with a norm pole (NH-E), as illustrated in Fig. 1.
In the NH-A class, all three Chern numbers vanish, and the right and left CLS bases are independently complete. In the NH-C class, the continuity of $\Pbr$ robustly preserves the locking $C_R=C_L=C_{\rm bi}$ at a common, potentially non-zero integer, establishing non-Hermitian critical topological flat bands. 
Once continuity is lost (NH-D), this topological locking collapses, rendering either $C_R$ or $C_L$ ill-defined. 
Finally, the NH-E class---a paradigm unique to non-Hermitian systems---features a geometric norm pole in $\Pbr$ that accommodates a quantized Chern-number mismatch ($C_R \neq C_L$). We demonstrate that this wave-function singularity is not a mere formal artifact but directly manifests as a massive, measurable enhancement in driven cross-orbital transport.

\begin{figure*}[t]
\includegraphics[width=180mm]{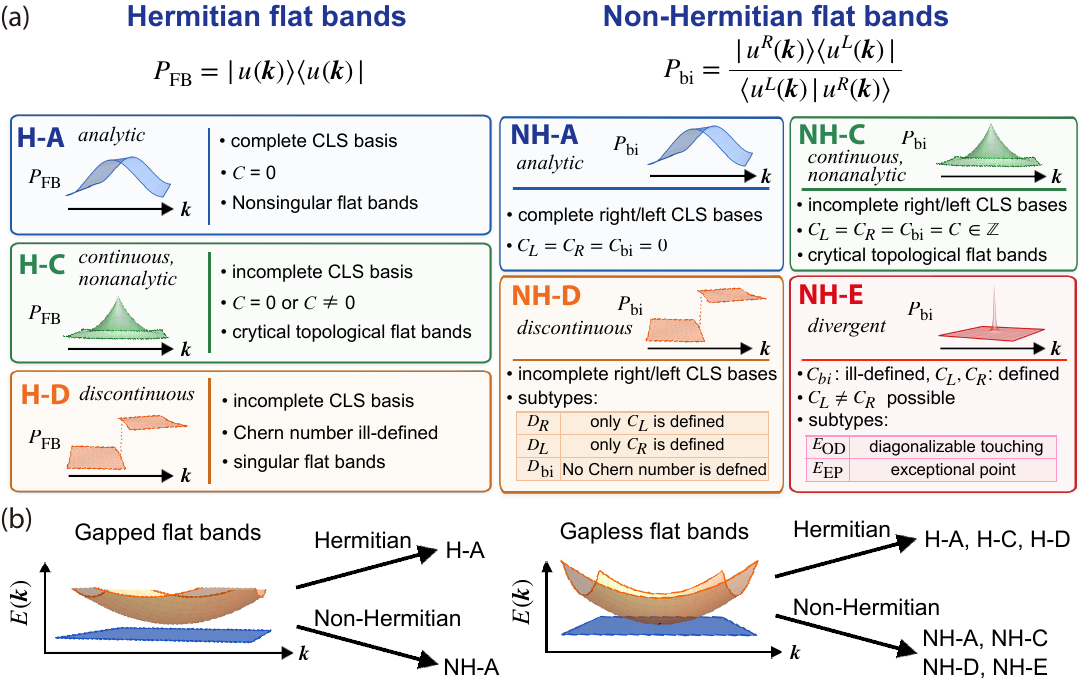} 
\caption{\textbf{Projector-regularity classification of Hermitian and non-Hermitian exact flat bands.}
(a) Hermitian flat bands are classified by the regularity of the flat-band projector $P_{\rm FB}(\bm k)$: analytic (H-A), continuous but nonanalytic (H-C), and discontinuous (H-D).
In non-Hermitian systems, the corresponding classification is governed by the biorthogonal projector $P_{\rm bi}(\bm k)$, yielding NH-A, NH-C, NH-D, and NH-E. (b) Relation to spectral isolation. A gapped flat band is necessarily analytic (H-A or NH-A), whereas gapless flat bands can realize all Hermitian classes or the NH-A--NH-E hierarchy, depending on the regularity of the corresponding projector.}\label{fig1}
\end{figure*}

\section{Setup}
We consider a finite-range tight-binding Hamiltonian
$H_{\alpha\beta}(\kk)=\sum_{\bm{R}\in S}t_{\alpha\beta}(\bm{R})\,e^{i\kk\cdot\bm{R}}$
with a finite hopping set $S$, $|S|<\infty$. Every matrix entry is a Laurent
polynomial in $e^{\pm ik_x},e^{\pm ik_y}$, so $H\in C^\omega(\mathbb{T}^2)$.
Shifting the exact flat-band energy to zero, flatness imposes
$\det H(\kk)\equiv 0$ for all $\kk\in\mathbb{T}^2$. Let the flat band is rank one, away from the touching at $\bm{k}_\star$.

In a Hermitian system, there exists a unique eigen-projector defined as $\PFB(\kk) = \lvert u(\kk) \rangle \langle u(\kk) \rvert$ for any normalized flat-band eigenvector $\lvert u(\kk) \rangle$. Since $\PFB^2 = \PFB = \PFB^\dagger$, the operator norm is $\|\PFB(\kk)\| = 1$. Consequently, a pole structure where the norm diverges cannot exist in Hermitian case.

In contrast, non-Hermiticity $H \neq H^\dagger$ fundamentally alters this picture. The right and left flat eigenvectors, defined by $H(\kk)\lvert u^R(\kk)\rangle=0$ and $\langle u^L(\kk)\rvert H(\kk)=0$, no longer coincide, causing the single projector to split into three:
\begin{align}
&P_L(\kk)=\frac{\lvert u^L(\kk)\rangle\langle u^L(\kk)\rvert}{\langle u^L(\kk)\vert u^L(\kk)\rangle},\quad
P_R(\kk)=\frac{\lvert u^R(\kk)\rangle\langle u^R(\kk)\rvert}{\langle u^R(\kk)\vert u^R(\kk)\rangle},\nonumber \\
&~~~~~~~~~~~~~~~~~~~~\Pbr(\kk)=\frac{\lvert u^R(\kk)\rangle\langle u^L(\kk)\rvert}{\langle u^L(\kk)\vert u^R(\kk)\rangle}.
\label{eq:three_projectors_formal}
\end{align}
By construction, the projectors $P_R(\kk)$ and $P_L(\kk)$ remain Hermitian orthogonal rank-one operators with a strict unit norm, $\|P_R(\kk)\|=\|P_L(\kk)\|=1$. The biorthogonal projector $\Pbr(\kk)$, by contrast, acts as an oblique idempotent whose norm is $\|\Pbr(\kk)\|
=
\frac{\|u^R(\kk)\|\,\|u^L(\kk)\|}
{|\langle u^L(\kk)|u^R(\kk)\rangle|}.$.
Throughout, we define $\chi(\bm k)$ as the left-right pairing, i.e., $\chi(\bm k)\equiv\langle u^L(\bm k)\vert u^R(\bm k)\rangle$.

Provided that the projector $P_{R}(\bm{k})$($P_{L}(\bm{k})$) extend continuously over the entire Brillouin torus, it define well-behaved right(left) complex line bundles with topological Chern number $C_R$($C_L$)~\cite{Kohmoto1985,Avron1983,ShenZhenFu2018}. For merely continuous projectors, these invariants are rigorously evaluated via the winding of transition functions between local frames, or equivalently through Wilson-loop or lattice-gauge constructions~\cite{Kohmoto1985,Alexandradinata2014,Fukui2005}. When $P_X(\bm{k})$ is additionally differentiable, each invariant admits the standard Berry-curvature expression~\cite{ShenZhenFu2018,Avron1983}:
\begin{equation}
C_X = \frac{1}{2\pi i} \int_{\rm BZ}d^2k\, \mathrm{Tr} \left[ P_X \left[ \partial_{k_x}P_X, \partial_{k_y}P_X \right] \right] \,, 
\label{eq:CRCL_formal}
\end{equation}
where $X=R,L$.
Likewise, on local patches where the right and left eigenvectors can be chosen continuously with the biorthogonal normalization $\langle u^L(\bm{k})\vert u^R(\bm{k})\rangle = 1$, their transition functions define the biorthogonal Chern number $C_{\rm bi}$~\cite{ShenZhenFu2018}. 

\section{Hermitian case}
We first review the classification of the flat band in Hermitian case~\cite{li2026stable}, which serves as the reference baseline for our non-Hermitian analysis.
Here, the right and left eigenvectors coincide, causing the three objects in Eq.~\eqref{eq:three_projectors_formal} to collapse into a single Hermitian projector $\PFB(\kk)$ with strict unit norm, $\|\PFB(\kk)\|=1$. 
The entire flat-band classification is then dictated by a single mathematical property---the regularity of $\PFB(\kk)$ at the touching set $\mathcal{K}$---which tripartitions the flat bands into three distinct regimes: $\PFB\in C^\omega$ (H-A), $\PFB\in C^0\setminus C^\omega$ (H-C), and $\PFB\notin C^0$ (H-D). 
This regularity concurrently fixes both their topological character and their completeness of CLSs. 

\textit{H-A (analytic, $\PFB\in C^\omega$):} A global non-vanishing analytic frame exists, meaning that the underlying line bundle is topologically trivial; the Chern number is well-defined and vanishes ($C=0$), representing the flat-band manifestation of the strict-locality no-go theorem~\cite{ChenMazaheriSeidelTang2014}. 
Furthermore because of the following strict equivalence (see SM)
\begin{equation}
\text{complete CLS basis}\ \Longleftrightarrow\ \PFB\in C^\omega(\BZ),
\label{eq:hermitian_anchor}
\end{equation}
the CLSs fully span the flat band, defining the class of nonsingular flat bands~\cite{RhimYang2019,RhimYang2021}.


\textit{H-C (continuous but non-analytic, $\PFB(\kk)\in C^0\setminus C^\omega$):} The projector remains continuous, ensuring that a complex line bundle and an integer Chern number are well-defined. 
Crucially, the Chern number can be either zero or non-zero. The flat bands with $C\neq 0$ are generally designated as critical topological flat bands~\cite{li2026stable,yang2025fractional}.

\textit{H-D (discontinuous, $\PFB\notin C^0$):} The projector $\PFB(\kk)$ admits no continuous extension across the touching set, as its local limit is inherently direction-dependent. Consequently, no global flat line bundle can be defined, and the Chern number becomes ill-defined. These are designated as singular flat bands~\cite{RhimYang2019,RhimYang2021}.

\section{Non-Hermitian classification}%
When non-Hermiticity is switched on ($H\neq H^\dagger$), each flat band falls
into one of four global classes---NH-A, NH-C, NH-D, NH-E---set by the regularity
of the projector $\Pbr(\bm k)$ at the touching set $\mathcal{K}$. 

\subsection{NH-A (analytic, $\Pbr \in C^\omega$)}
The projector $\Pbr$ remains analytic across the entire Brillouin zone, $\Pbr \in C^\omega(\mathbb{T}^2)$. Equivalently, there exist nowhere-vanishing Laurent-polynomial right and left flat-band eigenvectors $\lvert u^R(\bm{k}) \rangle$ and $\langle u^L(\bm{k}) \rvert$ with a nowhere-vanishing pairing $\chi(\bm k) \neq 0$ for all $\bm{k} \in \mathbb{T}^2$. Consequently, both projectors are analytic ($P_R, P_L \in C^\omega$), the right and left CLSs are each complete, and all topological invariants vanish $C_R = C_L = C_{\rm bi} = 0$ (see SM).
Crucially, the right and left CLS being separately compact and complete does not
make the biorthonormal resolution compact. The biorthonormal dual frame
$\langle \tilde{u}^L(\bm{k}) \rvert = \langle u^L(\bm{k}) \rvert / \chi(\bm{k})$
is compactly supported if and only if $\chi^{-1}$ is itself a Laurent polynomial (equivalently, $\chi$ is a monomial).

As examples, we consider two minimal NH-A models as
\begin{equation}
H_1=\begin{pmatrix}-ab & a^{2}\\[2pt] b^{2} & -ab\end{pmatrix},
\qquad
H_2=\begin{pmatrix}-b & a\\[2pt] 2b & -2a\end{pmatrix},
\end{equation}
where $a=e^{ik_x}$, $b=e^{ik_y}$.
They share the right flat eigenvector $\lvert u^R\rangle=(a,b)^{\mathsf T}$ and
differ only in the left eigenvector: $\langle u_1^L|=(b,a)$ for $H_1$ and $\langle u_2^L|=(2,1)$ for $H_2$. 
Both are strictly classified as NH-A since their projectors $\Pbr(\bm{k})=\mathbf{1}+H/\chi$ remain analytic across the entire Brillouin zone [$\Pbr \in C^\omega(\BZ)$]. Consequently, they both exhibit complete and compact right and left CLS bases with vanishing topological invariants, $C_R=C_L=C_{\rm bi}=0$.
The defining distinction lies exclusively in the algebraic structure of their pairings, which are given by the monomial $\chi_1=2ab$ for $H_1$ and the non-monomial $\chi_2=2a+b$ for $H_2$. Thus, the two models manifest a sharp contrast solely in the real-space compactness of their biorthogonal dual eigenvectors (see SM for full mathematical derivations).

\subsection{NH-C (continuous, $\Pbr \in C^0\setminus C^\omega$)}
The projector $\Pbr$ extends continuously across the entire Brillouin zone but
fails to be analytic at the touchings, $\Pbr \in C^0(\mathbb{T}^2)\setminus C^\omega$.
Continuity alone locks the three Chern numbers together (see SM),
\begin{equation}
\Pbr \in C^0(\BZ)\ \Longrightarrow\ C_R = C_L = C_{\rm bi}=C,
\label{eq:locking}
\end{equation}
and this common topological value can be non-zero.
Significantly, while standard non-Hermitian band theory guarantees $C_R=C_L=C_{\rm bi}$ only for isolated bands~\cite{ShenZhenFu2018}, Eq.~\eqref{eq:locking} establishes that the topological identity survives even when the band touches a different band, provided $\Pbr$ remains continuous at the touching.
When $C\neq 0$, the basis of CLSs becomes incomplete in stark contrast to the NH-A class, because a global analytic frame is fundamentally obstructed,.

To provide an realization of the NH-C class, we introduce $s=\sin k_x-i\sin k_y$, $p=2-\cos k_x-\cos k_y$, and define the spinor $\lvert q\rangle=(s,\,s+p)^{\mathsf T}$. We construct the Hamiltonian via the similarity transformation
\begin{equation}
H_\rho = S\,\lvert q\rangle\langle q\rvert\,S^{-1},
\end{equation}
where $ S=\mathrm{diag}(\rho,1)$ with $,\ 0<\rho\neq1$, and $\langle q\rvert$ denotes the conjugate transpose of $\lvert q\rangle$. 
Its spectrum are given as 
\begin{equation}
E_0(\bm{k})=0,\qquad E_d(\bm{k})=\lvert s\rvert^{2}+\lvert s+p\rvert^{2}.
\end{equation}
Thus the flat band touches the dispersive band exclusively at $\bm k =\bm 0$.
The zero-energy flat-band projector is
\begin{equation}
\Pbr(\kk)
=
\frac{1}{|s|^2+|s+p|^2}
\begin{pmatrix}
|s+p|^2 & -\rho\,s(s+p)^*\\
-\rho^{-1}(s+p)s^* & |s|^2
\end{pmatrix},
\label{eq:Hrho_flat_projector}
\end{equation}
and $\Pbr\in C^0\setminus C^\omega$. 
From Eq.~(\ref{eq:CRCL_formal}), $C_R=C_L=C_{\rm bi}=-1$. 

The same classification holds for a genuinely complex-spectrum model
\begin{align}H_\gamma=\lvert q\rangle\langle q\rvert\,T_\gamma, \qquad T_\gamma \equiv \begin{pmatrix}1&\gamma e^{ik_x}\\ 0&1\end{pmatrix}
\end{align}
where $0<\lvert\gamma\rvert<2$.
Its flat-band projector remains $\Pbr\in C^0\setminus C^\omega$ at the sole touching, so that $H_\gamma$ belongs to the NH-C class and carries the same locked Chern nuber, $C_R=C_L=C_{\rm bi}=-1$, as shown in SM.
From the non-Hermitian analogue of the $H-C$ class~\cite{li2026stable}, we refer NH-C flat bands with a nonzero locked Chern number as Non-Hermitian critical topological flat bands.

\subsection{NH-D (bounded but discontinuous, $\Pbr\notin C^0$)}
Near the touching set the projector $\Pbr$ stays bounded, $\lVert\Pbr(\bm k)\rVert<\infty$,
but admits no continuous extension: its limit at $\bm k_\star\in\mathcal{K}$ depends on the direction of approach. NH-D is thus separated from NH-C by the loss of continuity and from NH-E by the absence of a norm pole. 
More explicitly,  NH-D is classified into three subtypes based on the individual regularities of the projectors $P_R(\bm{k})$ and $P_L(\bm{k})$ as
\begin{equation}
\begin{array}{c|ccc}
& P_R & P_L & \text{well-defined}\\\hline
D_R & \text{disc.} & \text{cont.} & C_L\\
D_L & \text{cont.} & \text{disc.} & C_R\\
D_{RL}& \text{disc.} & \text{disc.} & \text{none}
\end{array}
\label{eq:NHD_subtypes}
\end{equation}
The biorthogonal $C_{\rm bi}$ is undefined throughout NH-D, since it requires a globally continuous $\Pbr$. 
This classification explicitly highlights the breakdown of the topological phase synchronization—the locking condition $C_R = C_L = C_{\mathrm{bi}}$ established in the continuous regime.

As an example we consider the following model. With $s=\sin k_x-i\sin k_y$,
$p=2-\cos k_x-\cos k_y$, set
\begin{equation}
\lvert q\rangle=\begin{pmatrix}s\\ s+p\end{pmatrix},\quad
\lvert q_\perp\rangle=\begin{pmatrix}-(s+p)^*\\ s^*\end{pmatrix},
\end{equation}
and, for $0<\lambda<1$, $\lvert v_\lambda\rangle=\lvert q\rangle+\lambda\lvert q_\perp\rangle$,
\begin{equation}
H^{D_L}_\lambda=\lvert v_\lambda\rangle\langle q\rvert .
\end{equation}
The spectrum consists of $E_{\rm FB}=0$ and $E_d=\lvert s\rvert^2+\lvert s+p\rvert^2$, and the flat band touches the
dispersive band only at the $\Gamma$ point ($\bm{k} = \bm{0}$). 
The right flat eigenvector
$\lvert u^R\rangle=\lvert q_\perp\rangle$ is direction-independent as $\bm k \to \Gamma$.
Consequently, the right projector is continuous ($P_R \in C^0$), yielding a well-defined Chern number $C_R = -1$. 
In contrast, the left flat eigenvector is instead
direction-dependent at $\Gamma$, so $P_L \notin C^0$, and hence $C_L$ is undefined.

\subsection{NH-E (divergent, $\lVert\Pbr\rVert\to\infty$)}
At the touching point $\kk_\star$ the projector $\Pbr$ becomes unbounded, $\lim_{\bm k\to \bm k_\star}\lVert\Pbr(\bm k)\rVert=\infty$.
By the Petermann identity $\lVert\Pbr\rVert=\lVert u^R\rVert\,\lVert u^L\rVert/\lvert\chi\rvert$,
this occurs precisely when the pairing $\chi$ vanishes faster than
$\lVert u^R\rVert\lVert u^L\rVert$. 
The pole arises from two distinct
mechanisms, distinguished by the
order of the resolvent pole $(\omega-H)^{-1}$
at the flat energy:
\begin{equation}
\begin{array}{c|cc}
& H(\bm k_\star) & \text{resolvent pole}\\\hline
E_{\rm OD} & \text{diagonalizable} & \omega^{-1}\\
E_{\rm EP} & \text{defective (Jordan, } m\ge2) & \omega^{-m}
\end{array}
\label{eq:NHE_subtypes}
\end{equation}
It is noteworthy that, in NH-E, $C_{\rm bi}$ is undefined similar to NH-D, but $C_R$ and $C_L$ are well defined, repectively. 

In $E_{\rm OD}$ (ordinary degeneracy) the touching is diagonalizable, so each band retains a simple pole and the divergence of $\Pbr$ originates from the vanishing $\chi(\bm k_\star)$ rather than from a Jordan block. 
When both projectors $P_R$ and $P_L$ extend continuously over the
entire Brillouin zone, the Chern numbers $C_R$ and $C_L$ are separately
well defined, and their mismatch is given by the total pairing-defect
charge,
\begin{equation}
C_R-C_L=\sum_j q_j,\qquad q_j=\frac{1}{2\pi}\oint_{\gamma_j}d\arg\chi ,
\label{eq:pairing_defect_charge}
\end{equation}
where $\gamma_j$ is a positively oriented loop enclosing the isolated
pairing zero $\bm{k}_j$.

As an example of $E_{\rm OD}$, we consider
\begin{equation}
H_{\rm OD}=\begin{pmatrix}p & -s\\ 0 & 0\end{pmatrix},
\end{equation}
where $s= \sin k_x - i \sin k_y$, $p=2-\cos k_x-\cos k_y$.
The eigenvectors and $\Pbr$ of the flat band near $\Gamma$ are given by 
\begin{align}
    \quad
\lvert u^R\rangle=\begin{pmatrix}1\\ p/s\end{pmatrix},~
\langle u^L\rvert=(0,\ 1),~
\Pbr=\begin{pmatrix}0 & s/p\\ 0 & 1\end{pmatrix},
\end{align}
with $\chi=p/s$. As $\bm k\to \Gamma$, $s/p\to\infty$ is a pole, while the touching $H_{\rm OD}(\Gamma)=0$ is diagonalizable. A straightforward calculation leads to $C_R=1$ and $C_L=0$.

In $E_{\rm EP}$ (exceptional point) the touching is no longer an ordinary band
touching, and $H(\bm
k_\star)$ can no longer be diagonalized. More explicitly, at a touching point $H(\bm k_\star)\equiv N$ has both
eigenvalues equal to the flat energy but $N\neq0$; a single
matrix with these properties is nilpotent. When $N^{m}=0$ and $N^{m-1}\neq 0$ are satisfied, the generalized eigenspace contains a Jordan block of size \(m\). Because the right and left eigenvectors merge, their overlap
collapses---the self-orthogonality $\chi\to0$ is forced by the defect itself, not by
an accidental cancellation---and $\lVert \Pbr \rVert$ diverges.

Similar to $E_{\rm OD}$, it can exhibit $C_R\neq C_L$. 
However, this class leaves two operational fingerprints absent from $E_{\rm OD}$. First, the
resolvent develops a higher-order pole. For instace, using $N^{2}=0$,
\begin{equation}
(\omega\mathbf{1}-N)^{-1}=\frac{\mathbf{1}}{\omega}+\frac{N}{\omega^{2}},
\end{equation}
so the flat energy is a double pole ($\omega^{-2}$), in contrast to the simple
poles of a diagonalizable touching ($E_{\mathrm{OD}}$). Second, the corresponding time evolution is no
longer purely oscillatory but grows polynomially,
\begin{equation}
e^{-iNt}=\mathbf{1}-iNt ,
\end{equation}
a linear-in-$t$ transient that is the dynamical signature of the Jordan block. More
generally a size-$m$ block produces an $\omega^{-m}$ pole and a degree-$(m{-}1)$ polynomial in $t$.

An example is given by
\begin{equation}
H_{\rm EP}=\begin{pmatrix}1 & 1\\ p-1 & p-1\end{pmatrix},\qquad p=2-\cos k_x-\cos k_y,
\end{equation}
which has eigenvalues $0$ and $p$. The flat-band
projector is $\Pbr=\mathbf{1}-H_{\rm EP}/p$ and diverges as $\lVert\Pbr\rVert\sim\lvert\bm k\rvert^{-2}$ near $\Gamma$. 
At the touching itself the Hamiltonian becomes
$H_{\rm EP}(\Gamma)\equiv N\neq 0$, which is nonzero but obeys $N^{2}=0$. Although this model exhibits $C_R=C_L=0$, NH-E$_{\rm EP}$ can have different $C_R$ from $C_L$.

Crucially, $H_{\rm EP}$ displays a phenomenon with no Hermitian analogue. Its right
and left flat eigenvectors ($\lvert u^R\rangle=(-1,~1 )^T$, $\langle u^L\rvert=(1-p,\ 1)$)
are both nowhere-vanishing Laurent polynomials, so the right and left CLS are each complete. However, the pairing $\chi=p$ vanishes at $\Gamma$, so the biorthonormal dual $\langle u^L\rvert/\chi$ has a pole. Thus, completeness of the right and left CLSs does not guarantee a bounded $\Pbr$ in sharp contrast to the Hermitian case.

\section{signatures of Chern number mismatch in cross-orbital transfer.}
A nonzero Chern number mismatch ($C_R \neq C_L$) topologically guarantees a norm pole in $\Pbr$.
This geometric norm pole is not a mere abstraction of Bloch eigenvectors, but manifests as a directly measurable anomaly in driven interorbital dynamics. 

Consider a two-orbital unit cell spanned by orbitals \(\lvert A\rangle\) and \(\lvert B\rangle\), such as sublattice sites
or coupled resonators. 
Under a monochromatic coherent drive on
$\lvert B\rangle$ with phenomenological damping rate $\eta>0$, the cross-orbital transfer intensity at the flat-band resonance ($\omega=0$) is
\begin{equation}
\mathcal{T}_{B\to A}(\bm k)=\bigl\lvert[G_\eta(0,\bm k)]_{AB}\bigr\rvert^2,
\end{equation}
where $G_\eta(\omega,\bm k)=[(\omega+i\eta)\mathbf{1}-H]^{-1}$.
Because the rank-one flat band dictates $H_{AB}=-E_d(\bm k)\,(\Pbr)_{AB}$, a straightforward calculation (SM) yields
\begin{equation}
\mathcal{T}_{B\to A}(\bm k)=\frac{E_d(\bm k)^2\,\lvert(\Pbr)_{AB}(\bm k)\rvert^2}
{\eta^2\,[\eta^2+E_d(\bm k)^2]},
\label{eq:transfer_compact}
\end{equation}
where $E_d(\bm k)\in \mathbb{R}$ indicates the dispersive band.
We characterize the strongest momentum-resolved response by $\mathcal{T}^{\max}\equiv \max_{\bm{k}\in\BZ} \mathcal{T}_{B\to A}(\bm{k})$. 
For any bounded projector, provided that $(\Pbr)_{AB}(\bm k_0)\neq0$ at some momentum $\bm k_0$ with $E_d(\bm k_0)\neq0$, Eq.~\eqref{eq:transfer_compact} yields the ordinary Lorentzian intensity scaling $\mathcal{T}^{\max}\propto \eta^{-2}$.
By contrast, a Chern mismatch induced pole in $\Pbr$ drives the anomalous scaling (see SM).
The reason is that the left and right eigenvectors play different roles in a non-Hermitian response: the left eigenvector selects how a drive on $\lvert B\rangle$ excites the flat-band mode, whereas the right eigenvector determines how that excited mode appears on $\lvert A\rangle$.
A Chern mismatch forces $\chi(\bm k)$ to vanish at the touching, so this input--output conversion acquires a singular geometric amplification. 
As a result, the ordinary Lorentzian resonance is enhanced by the projector pole, leading to an anomalous damping dependence of the maximal transfer.

\begin{figure}[t]
\includegraphics[width=90mm]{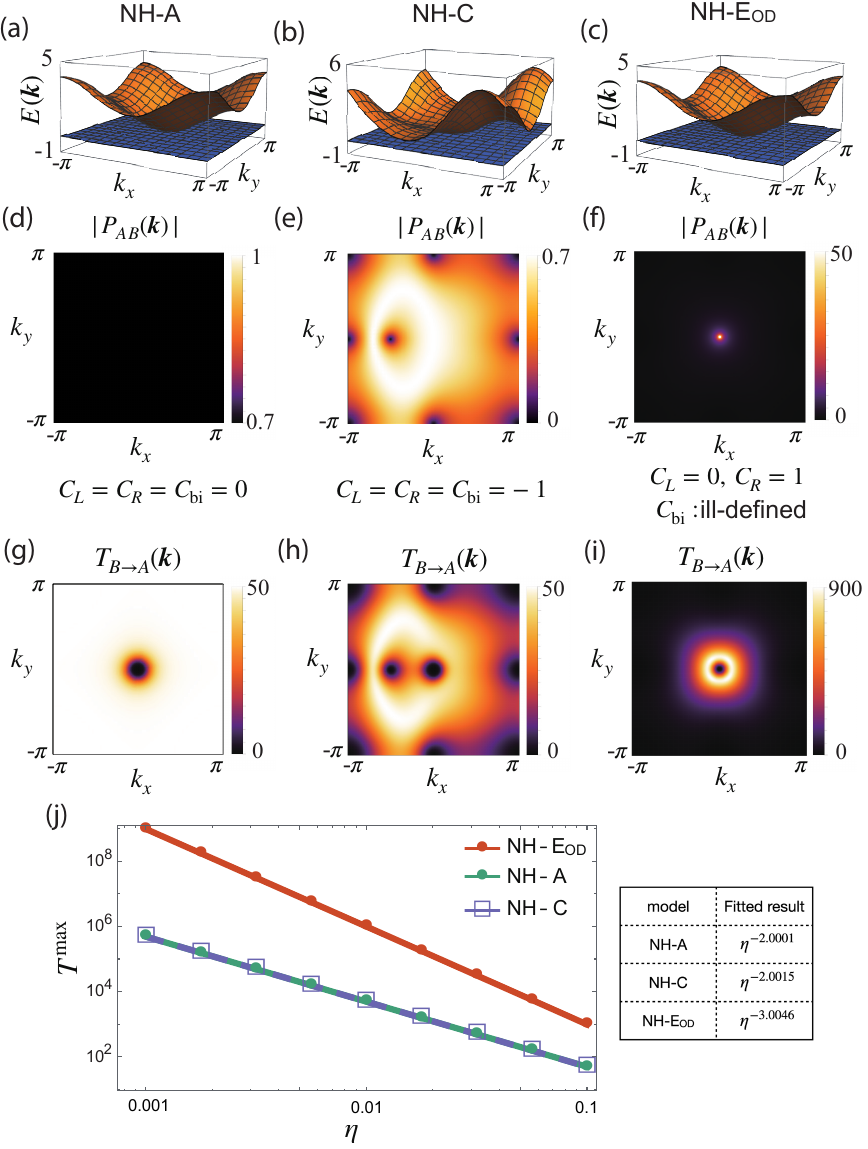} 
\caption{\textbf{Chern-mismatch-induced enhancement of driven
interorbital transfer.}
(a)--(c) Band structures of the NH-A, NH-C, and NH-E$_{\rm OD}$ control models (see SM for details). 
(d)--(f) Magnitude of the off-diagonal biorthogonal flat-band
projector element, $\lvert(\Pbr){AB}(\bm k)\rvert$. $C_L$, $C_R$ and $C_{\rm bi}$ indicate left-eigenstate Chern number, right-eigenstate Chern number and biorthogonal Chern number, respectively.
(g)--(i) Cross-orbital transfer intensity $\mathcal{T}_{B\to A}(\bm k)$ at $\eta=0.1$. 
(j) The maximum transfer intensity $\mathcal{T}^{\max}$ as a function of damping rate $\eta$ and the fitted exponents.}\label{fig2}
\end{figure}

As an example, we compare three control models from NH-A, NH-C, and NH-E$_{\rm OD}$, engineered to share the same flat-band energy and the identical leading dispersion $E_d(\bm k)\simeq\lvert\bm k\rvert^2/2$ near $\Gamma$ (see SM and Fig.~\ref{fig2}). 
For the regular NH-A and NH-C models, the biorthogonal projector $\Pbr$ remains strictly bounded, as shown in Fig.~\ref{fig2}(d,e), meaning the interorbital element of the Hamiltonian simply tracks the dispersion as $H_{AB} = -E_d(\Pbr)_{AB} \sim O(\lvert\bm{k}\rvert^2)$.
In contrast, for the Chern-mismatched NH-E$_{\rm OD}$, the topological mismatch forces a geometric pole in the biorthigonal projector as $(\Pbr)_{AB}\sim\lvert\bm{k}\rvert^{-1}$, as shown in Fig.~\ref{fig2}(f), which leads to $H_{AB}\sim O(\lvert\bm k\rvert)$. 
This geometric conversion culminates in a qualitatively distinct observable profile (Fig.~\ref{fig2}(g-j)): the bounded NH-A and NH-C configurations exhibit the ordinary scaling $\mathcal{T}_{\rm A/C}^{\max}\sim\eta^{-2}$, whereas the Chern-mismatched NH-E$_{\rm OD}$ model exhibits an enhanced scaling of $\mathcal{T}_{\rm OD}^{\max}\sim\eta^{-3}$.

\section{Conclusion}
We have established a projector-based classification of non-Hermitian flat bands.
The relevant object is the biorthogonal flat band projector
$\Pbr$, whose regularity at band touching yields four distinct classes: analytic (NH-A), continuous but nonanalytic (NH-C), bounded but discontinuous (NH-D), and divergent (NH-E).
This hierarchy unifies several properties that are disconnected in an energy-spectrum-based description.
In NH-A, right and left CLSs are separately complete and all Chern numbers vanish, although compactness of the biorthogonal dual frame requires the additional algebraic condition that $\chi^{-1}$ be Laurent-polynomial.
In NH-C, continuity of $\Pbr$ protects a global biorthogonal line bundle and enforces $C_R=C_L=C_{\rm bi}$, defining non-Hermitian critical topological flat bands.
Once continuity is lost in NH-D, the Chern number locking breaks down, and $C_R$ or $C_L$ is ill-defined.
In NH-E, the vanishing left-right pairing produces a pole in $\Pbr$, arising either at a diagonalizable touching or at an exceptional point, the latter additionally characterized by higher-order resolvent poles and polynomial temporal evolution.
Furthermore, in NH-E class, Chern number mismatch $C_R\neq C_L$ can occur. We showed that this singularity has a direct observable manifestation in resonant cross-orbital transfer.

Several directions follow naturally from the present work.
First, the present classification assumes a rank-one flat band.
Its extension to multiband flat manifolds should involve non-Abelian oblique projectors, where the interplay between exceptional degeneracies, Wilson-loop topology, and compact localized frames may be substantially richer.
Second, the phenomenon of a topological Chern mismatch ($C_R \neq C_L$) is fundamentally not restricted to flat-band architectures. In any general non-Hermitian system having a pole in a projector, the decoupling of right and left eigenrays can stabilize a non-zero topological mismatch. Generalizing our geometric framework to these purely dispersive settings would unveil how right-left wave-function mismatches govern non-normal transport.
Third, incorporating many-body interactions or collective instabilities into these singular non-Hermitian bands represents a crucial next step. While Hermitian flat bands are renowned for harboring strongly correlated phenomena—ranging from fractional Chern insulators to unconventional superconductivity and magnetic ordering— the introduction of non-Hermiticity adds an uncharted physics. 

\begin{acknowledgments}
The authors thank Z. Song for useful discussions.
C.Oh acknowledge the support by Japan Society for
the Promotion of Science (JSPS), KAKENHI Grant No. JP25KF0186.
J.W.R. was supported by the National Research Foundation of Korea (NRF) Grant funded by the Korean government (MSIT) (grant no. RS-2023-NR068116) and the Ministry of Education (grant no. RS-2023-00285390).
\end{acknowledgments}

\bibliography{ref}

\end{document}